\documentstyle[11pt]{article}

\begin{document}

\begin{center}
{\Large {\bf Quantum probability distribution of arrival times and
probability current density \vspace*{0.6cm}\\}} {\large {V. Delgado 
\vspace*{.2cm}}}\\{\it Departamento de F\'\i sica Fundamental y
Experimental, \\Universidad de La Laguna, 38205-La Laguna, Tenerife, Spain\\%
e-mail: vdelgado@ull.es\\{\rm (May 19, 1998)}\vspace*{0.7cm}\\}
\end{center}

\begin{abstract}
This paper compares the proposal made in previous papers for a quantum
probability distribution of the time of arrival at a certain point with the
corresponding proposal based on the probability current density.
Quantitative differences between the two formulations are examined
analytically and numerically with the aim of establishing conditions under
which the proposals might be tested by experiment. It is found that quantum
regime conditions produce the biggest differences between the formulations
which are otherwise near indistinguishable. These results indicate that in
order to discriminate conclusively among the different alternatives, the
corresponding experimental test should be performed in the quantum regime
and with sufficiently high resolution so as to resolve small quantum 
effects.
\end{abstract}

\hspace{6. pt} PACS number(s): 03.65.Bz, 03.65.Ca

\vspace{1.4 cm}

\begin{center}
{\large {\bf {I. INTRODUCTION\vspace{.1 cm}\\}}}
\end{center}

The problem of incorporating the time-of-arrival concept in the theory of
quantum measurement has remained controversial over the years, and even
nowadays this question is open to debate [\ref{Pauli}--\ref{Hall}]. In
recent times this issue has acquired renewed interest in part due to the
development of new experimental techniques for probing quantum systems in
the time domain. For instance, by exciting an atomic system with a pulsed
laser and measuring the subsequent flux of electrons ejected from
autoionizing states, as a function of the time of arrival at the detector,
one can gain important physical information which is not obtainable by
probing the system in the more familiar energy domain [\ref{frobi}]. On the
other hand, the time domain is more related to the macroscopic phenomena and
for this reason turns out to be particularly suitable for investigating
quantum systems at the mesoscopic scale [\ref{onofrio}].

Another related issue that has stimulated considerable theoretical effort is
that concerning the definition and characterization of tunneling times [\ref
{Lan},\ref{Rev}]. In connection with this problem,  Dumont and Marchioro
proposed the probability current density as a quantum definition for the
(unnormalized) probability distribution of arrival times at an asymptotic
point behind a one-dimensional potential barrier [\ref{Du}].

There exists additional motivation for trying to incorporate such a
definition into the formalism of quantum mechanics. Indeed, the average
current $\langle J(X)\rangle $ of a classical statistical ensemble of
particles propagating in one spatial dimension along a well-defined
direction plays the role of a probability distribution of arrival times at $X
$. A simple way for translating such a result into the framework of quantum
mechanics consists in invoking the Weyl-Wigner quantization rule, which
provides a prescription for constructing a quantum operator $\hat A(\hat
X,\hat P)$ corresponding to a given classical dynamical variable $A(x,p)$ [%
\ref{Wigner2},\ref{Cohen}],

\begin{equation}
\label{epg1}A(x,p)\rightarrow \hat A(\hat X,\hat P)=\frac 1{4\pi ^2}\!\int
\!\!\int \!\!\int \!\!\int \!A(x,p)e^{i[\theta (\hat X-x)+\tau (\hat
P-p)]}\,dx\,dp\,d\theta \,d\tau .
\end{equation}
Furthermore, the operator so obtained has the nice property that its
expectation value is given by the classical expression

\begin{equation}
\label{unms1}\langle \hat A(\hat X,\hat P)\rangle =\!\int \!\!\int \!f_{{\rm %
W}}(x,p)\,A(x,p)\,dx\,dp, 
\end{equation}
with the {\em Wigner function} $f_{{\rm W}}(x,p)$ playing the role of a
quasiprobability distribution function in phase space.

The Weyl-Wigner quantization rule must be used with caution for it does not
necessarily lead to the correct quantum operator. In the present context,
one obtains that the Weyl-Wigner operator corresponding to the classical
current $J(X)=p/m\,\delta (x-X)$ is nothing but the usual current operator

\begin{equation}
\label{curop}\hat J(X)=\frac 1{2m}\left( \hat P\,|X\rangle \langle
X|+|X\rangle \langle X|\,\hat P\right) .
\end{equation}
However, unlike the classical case, because of the fact that $\hat J(X)$ is
not positive definite, its expectation value cannot be properly considered
as a probability distribution of arrival times. It has been argued,
nonetheless, that asymptotically far from a potential barrier the
transmitted current becomes positive,  and this circumstance justifies its
interpretation as a probability distribution [\ref{Du},\ref{Mug}]. In this
regard, McKinnon and Leavens [\ref{Leav3},\ref{NEW3}] have also shown that
within the framework of Bohmian mechanics it is possible to unambiguously
define a probability distribution of the time of arrival in terms of the
modulus of the probability current density. However, even though such a
definition circumvents the problem mentioned above, in principle there is no
justification for extrapolating it to the framework of standard quantum
mechanics.

A natural way for introducing time into the quantum framework as a physical
variable consists in considering it as such already at the classical level
(a fact that can be implemented by making a suitable canonical
transformation) and then quantizing the corresponding formulation by using
the canonical quantization method [\ref{QF}] in order to look for the
desired probability distribution in terms of the spectral decomposition of
an appropriate self-adjoint operator. In doing so, one arrives at a time
operator defined as the operator canonically conjugate to the relevant
Hamiltonian [\ref{VDB4},\ref{Raz},\ref{Ko}]. However, in general, no such a
self-adjoint operator exists [\ref{Pauli},\ref{Allco},\ref{VDB4}]. This is
the technical reason that explains to a great extent the difficulty found
for incorporating a time operator into the quantum formalism.

A reasonable way of circumventing this problem consists in looking instead
for a self-adjoint operator with dimensions of time not strictly conjugate
to the Hamiltonian. Even though there exist  appreciable differences among
them, the approaches of Kijowski [\ref{Kij}], Grot {\em et al.} [\ref{Grot}%
], as well as the one developed in Refs. [\ref{VDB4}] and [\ref{VDB5}] can
be ascribed to this  category. The first two approaches are concerned with
the time of arrival of a free particle, and its supposed range of validity
includes quantum states having, in the momentum representation, positive-
and negative-momentum components, while the latter is also applicable
(asymptotically) in the presence of a one-dimensional scattering potential
and its range of validity is restricted to quantum states having either
positive- or negative-momentum contributions. In this paper we shall focus
on this latter approach. It should be remarked, however, that within their
common range of applicability all of  them provide the same theoretical
prediction for the probability distribution of the time of arrival at a
certain point.

Agreement with a conclusive experimental test is the ultimate requirement
for establishing the validity of any theoretical proposal. However,
discriminating experimentally among different alternatives is not always a
straightforward matter. It may happen that under certain experimental
conditions predictions corresponding to different proposals become
indistinguishable in practice. This is the case in the present context when
considering quantum states largely semiclassical in character. Indeed, in
the semiclassical limit [\ref{VDB5}] the proposal for the probability
distribution of arrival times based on the operator approach coincides with
that based on the modulus of the probability current density, which is the
result obtained by McKinnon and Leavens within Bohmian mechanics [\ref{Leav3}%
,\ref{NEW3}]. More generally, since in this limit the quantum current
becomes necessarily positive, it follows that the predictions based on the
operator approach become in fact  indistinguishable from those based in
general on the probability current density. Consequently, any experimental
test performed under these particular conditions would be inconclusive. It
is therefore worthwhile to investigate quantitatively to what extent
appreciable differences among the competing proposals can be expected as
well as to examine how such differences depend on the various controllable
parameters. This is the main purpose of the present work [\ref{Nota}]. More
specifically, quantitative differences between the two formulations will be
examined analytically and numerically (as a function of both the initial
quantum state describing the particle and the parameters characterizing an
intermediate potential barrier) with the aim of establishing conditions
under which the proposals might be tested by experiment. To this end we
shall begin by briefly reviewing the required formulation.

\vspace{1.4 cm}

\begin{center}
{\large {\bf {II. PROBABILITY DISTRIBUTION OF ARRIVAL TIMES\vspace{.1 cm}\\}}%
}
\end{center}

Consider a quantum particle moving along the $x$ axis toward a detector
located at a certain asymptotic point $X$ behind a one-dimensional
scattering center $V(\hat X)$. In looking for a probability distribution of
the time of arrival for such a physical system, we introduced in previous
papers [\ref{VDB4},\ref{VDB5}] a self-adjoint operator with dimensions of
time $\hat {{\cal T}}(X)$ defined in terms of its orthogonal spectral
decomposition by

\begin{equation}
\label{ec68}\hat {{\cal T}}(X)=\!\int_{-\infty }^{+\infty }\!d\tau \,\tau
\,|\tau ;X\rangle \langle \tau ;X|, 
\end{equation}

\begin{equation}
\label{indp}|\tau ;X\rangle \equiv e^{i\,{\rm sgn}(\hat P\,)\frac{\hat P\,^2%
}{2m}\tau /\hbar }\,\sqrt{\frac{|\hat P\,|}m}\,|X\rangle . 
\end{equation}
The operators ${\rm sgn}(\hat P)$ and $\sqrt{|\hat P\,|}$ are in turn given
by the expressions

\begin{equation}
\label{hbm}\sqrt{|\hat P\,|}\equiv \int_{-\infty }^{+\infty }dp\sqrt{|p\,|}%
\,|p\rangle \langle p|, 
\end{equation}

\begin{equation}
\label{sgdf}{\rm sgn}(\hat P)\equiv \!\int_0^\infty dp\left( |p\rangle
\langle p|-\mid \!-p\rangle \langle -p\!\mid \right) , 
\end{equation}
where the momentum eigenstates $\left\{ |p\rangle \right\} $ are assumed to
be normalized as $\langle p|p^{\prime }\rangle =\delta (p-p^{\prime })$.

Note that the above equations define, in fact, a one-parameter family $%
\{\hat {{\cal T}}(X)\}$ of self-adjoint operators (labeled by the position $X
$ of the detector) which are canonically conjugate to the operator $\hat {%
{\cal H}}\equiv {\rm sgn}(\hat P)\,\hat H_0$, with $\hat H_0\equiv \hat
P^2/2m$ being the energy of the free particle.

Let $\Theta (+\hat P)$ [$\Theta (-\hat P)$] represent the projector onto the
subspace spanned by plane waves with positive [negative] momenta,

\begin{equation}
\label{ec16}\Theta (\pm \hat P)=\!\int_0^\infty dp\mid \!\pm p\rangle
\langle \pm p\!\mid . 
\end{equation}
By taking advantage of the resolution of the unity $\Theta (+\hat P)+\Theta
(-\hat P)={\bf \hat 1}$, we can rewrite the eigenstates $|\tau ;X\rangle $
(which are manifestly symmetric under time reversal) in the form 
\begin{equation}
\label{indp2n}|\tau ;X\rangle =\Theta (+\hat P)\,e^{i\hat H_0\tau /\hbar }\,%
\sqrt{\frac{|\hat P\,|}m}\,|X\rangle +\Theta (-\hat P)\,e^{-i\hat H_0\tau
/\hbar }\,\sqrt{\frac{|\hat P\,|}m}\,|X\rangle , 
\end{equation}
which involves the state $\sqrt{|\hat P\,|/m}\,|X\rangle $ translated
(freely) both forward and backward in time by the amount $\tau $.

Substituting then Eq. (\ref{indp2n}) into Eq. (\ref{ec68}), one obtains

\begin{equation}
\label{eqt}\hat {{\cal T}}(X)=\!\Theta (+\hat P)\left[ \int_{-\infty
}^{+\infty }\!d\tau \,\tau \,\hat J_{{\rm I}}^{(+)}(X,\tau )\right] \Theta
(+\hat P)-\,\!\Theta (-\hat P)\left[ \int_{-\infty }^{+\infty }\!d\tau
\,\tau \,\hat J_{{\rm I}}^{(+)}(X,\tau )\right] \Theta (-\hat P), 
\end{equation}
where the positive-definite current $\hat J_{{\rm I}}^{(+)}(X,\tau )$ is a
straightforward quantum version (in the interaction picture) of the modulus
of the classical current $|J(X)|=|p|/m\,\delta (x-X)$,

\begin{equation}
\label{jph}\hat J_{{\rm I}}^{(+)}(X,\tau )=e^{i\hat H_0\tau /\hbar }\,\hat
J^{(+)}(X)\,e^{-i\hat H_0\tau /\hbar }, 
\end{equation}

\begin{equation}
\label{corr1}\hat J^{(+)}(X)\equiv \sqrt{\frac{|\hat P\,|}m}\,\delta (\hat
X\,-X)\,\sqrt{\frac{|\hat P\,|}m}. 
\end{equation}

Even though $\hat {{\cal T}}(X)$ is symmetric under time reversal, its
restrictions to the subspaces spanned by either positive- or
negative-momentum plane waves are not. This fact enables us to define a
probability distribution of the time of arrival for quantum states belonging
to either of such subspaces. To be specific, let us assume the particle
under study to be incident from the left of the potential barrier, and let
the state vector $|\psi _{{\rm in}}\rangle $ [which is assumed to satisfy
the identity $|\psi _{{\rm in}}\rangle \equiv \Theta (\hat P)\,|\psi _{{\rm %
in}}\rangle $]  represent the incoming asymptote of the actual scattering
state of the particle at $t=0$. The mean arrival time at an asymptotic point 
$X$ can then be defined consistently as [\ref{VDB4},\ref{VDB5}]

\begin{equation}
\label{ec77x}\langle t_X\rangle =\frac{\langle \psi _{{\rm tr}}|\hat {{\cal T%
}}(X)|\psi _{{\rm tr}}\rangle }{\langle \psi _{{\rm tr}}|\psi _{{\rm tr}%
}\rangle }=\frac 1{\langle \psi _{{\rm tr}}|\psi _{{\rm tr}}\rangle
}\!\int_{-\infty }^{+\infty }\!dt\,t\,\langle \psi _{{\rm tr}}|\hat J_{{\rm I%
}}^{(+)}(X,t)|\psi _{{\rm tr}}\rangle , 
\end{equation}
where $|\psi _{{\rm tr}}\rangle $ is the projection of the outgoing
asymptote (at $t=0$) onto the channel of transmitted particles, i.e.,

\begin{equation}
\label{ptrix}|\psi _{{\rm tr}}\rangle =\Theta (\hat P)|\psi _{{\rm out}%
}\rangle =\Theta (\hat P)\,{\hat S}\,|\psi _{{\rm in}}\rangle
=\!\int_0^\infty \!dp\,T(p)\,\langle p|\psi _{{\rm in}}\rangle \,|p\rangle , 
\end{equation}
with ${\hat S}$ and $T(p)$ being, respectively, the scattering operator and
the transmission coefficient characterizing the potential barrier.

It is worth noting the remarkable formal analogy between Eq. (\ref{ec77x})
and its corresponding classical counterpart. Indeed, the positive-definite
current $\langle \psi _{{\rm tr}}|\hat J_{{\rm I}}^{(+)}(X,t)|\psi _{{\rm tr}%
}\rangle $ enters the expression for $\langle t_X\rangle $ playing the role
of an (unnormalized) probability distribution. We can thus define the
probability  distribution of the time of arrival at the asymptotic point $X$
as

\begin{equation}
\label{pqua}P_X(t)\equiv \frac 1{{\sf T}}|\langle t;X|\psi _{{\rm tr}%
}\rangle |^2=\frac 1{{\sf T}}\langle \psi _{{\rm tr}}(t)|\hat
J^{(+)}(X)|\psi _{{\rm tr}}(t)\rangle , 
\end{equation}
where ${\sf T\equiv }\langle \psi _{{\rm tr}}|\psi _{{\rm tr}}\rangle
=\langle \psi _{{\rm tr}}(t)|\psi _{{\rm tr}}(t)\rangle $ is the
transmittance and we have written the latter expression in the more familiar
Schr\"odinger picture by introducing the (Schr\"odinger) freely evolving
transmitted state

\begin{equation}
\label{ecita}|\psi _{{\rm tr}}(t)\rangle \equiv e^{-i\hat H_0t/\hbar }|\psi
_{{\rm tr}}\rangle . 
\end{equation}

Equation (\ref{pqua}) along with Eqs. (\ref{ptrix}) and (\ref{ecita}) enable
us to compute the  desired probability distribution in terms of the ingoing
asymptote $|\psi _{{\rm in}}\rangle $. It is worth, however, obtaining an
alternative formula in terms of the actual scattering state $|\psi
(t\!=\!0)\rangle $. This can be accomplished by means of the M\o ller
operators,

\begin{equation}
\label{mol}\hat \Omega _{\pm }=\lim _{t\rightarrow \mp \infty }e^{i\hat
Ht/\hbar }\,e^{-i\hat H_0t/\hbar }, 
\end{equation}
which map the ingoing and outgoing asymptotic states onto the corresponding
scattering state

\begin{equation}
\label{rela}|\psi (t\!=\!0)\rangle =\hat \Omega _{+}|\psi _{{\rm in}}\rangle
=\hat \Omega _{-}|\psi _{{\rm out}}\rangle . 
\end{equation}
Using these relations and introducing the projector

\begin{equation}
\label{botc}{\hat {{\cal P}}}\equiv \hat \Omega _{-}\Theta (\hat P)\hat
\Omega _{-}^{\dagger }
\end{equation}
(which selects that part of a given state vector that will be transmitted),
one finally obtains (Appendix A)

\begin{equation}
\label{dxnt}P_X(t)=\frac 1{{\sf T}}\langle \psi (0)|{\hat {{\cal P}}\,}\hat
\Omega _{-}\hat J_{{\rm I}}^{(+)}(X,t)\hat \Omega _{-}^{\dagger }{\hat {%
{\cal P}}}|\psi (0)\rangle =\frac 1{{\sf T}}\langle \psi (t)|{\hat {{\cal P}%
}\,}\hat \Omega _{-}\hat J^{(+)}(X)\hat \Omega _{-}^{\dagger }{\hat {{\cal P}%
}}|\psi (t)\rangle ,
\end{equation}
where $|\psi (t)\rangle \equiv e^{-i\hat Ht/\hbar }|\psi (0)\rangle $ is the
usual Schr\"odinger state vector. It is interesting to note that the above
equation is merely the expectation value of the modulus of the current $\hat
J^{(+)}(X)$ in the quantum state $1/\sqrt{{\sf T}}\,\hat \Omega
_{-}^{\dagger }{\hat {{\cal P}}}|\psi (t)\rangle $, which, in turn, is the
normalized outgoing asymptote corresponding to that part of $|\psi
(t)\rangle $ that is going to be transmitted in the future.

In practice, whenever the actual scattering state at $t=0$ does not overlap
appreciably with the potential barrier, the state vectors $|\psi _{{\rm in}%
}\rangle $ and $|\psi (0)\rangle $ become physically indistinguishable [\ref
{Taylor}] and, consequently, one can legitimately use Eqs. (\ref{ptrix})--(%
\ref{ecita}) with the substitution $|\psi _{{\rm in}}\rangle \rightarrow $ $%
|\psi (0)\rangle $.

For our purposes it is convenient to write the expectation values of $\hat
J(X)$ and $\hat J^{(+)}(X)$ as

\begin{equation}
\label{prim}\langle \psi _{{\rm tr}}(t)|\hat J(X)|\psi _{{\rm tr}}(t)\rangle
=\frac 1{mh}\frac 12\left( I^{*}[p]I[1]+{\rm c.c.}\right) , 
\end{equation}

\begin{equation}
\label{segu}\langle \psi _{{\rm tr}}(t)|\hat J^{(+)}(X)|\psi _{{\rm tr}%
}(t)\rangle =\frac 1{mh}\left( I^{*}[\sqrt{p}]I[\sqrt{p}]\right) , 
\end{equation}
where we have introduced the functional

\begin{equation}
\label{funt}I\left[ f\right] \equiv \!\int_0^\infty
\!\!dp\,T(p)\,f(p)\,\langle p|\psi _{{\rm in}}\rangle \,e^{-i\frac{p^2}{2m}%
t/\hbar }e^{ipX/\hbar }. 
\end{equation}

Note finally that the free case is a particular case of the above
formulation with the M\o ller operators $\hat \Omega _{\pm }$ reducing to
the unit operator and, consequently, ${\hat {{\cal P}}}\rightarrow \Theta
(\hat P)$. Since the scattering operator can be written as ${\hat S=}\hat
\Omega _{-}^{\dagger }\hat \Omega _{+}$, it also follows that ${\hat
S\rightarrow }{\bf \hat 1}$, and hence, by virtue of Eq. (\ref{ptrix}), $%
T(p)\rightarrow 1$ and $|\psi _{{\rm tr}}\rangle \rightarrow |\psi _{{\rm in}%
}\rangle \rightarrow |\psi (0)\rangle $. With these substitutions the above
formulas are applicable to the study of the arrival time of a free particle
at a point $X$.

\vspace{1.4 cm}

\begin{center}
{\large {\bf {III. ANALYTICAL APPROXIMATION FOR THE EXPECTATION VALUE OF $%
\hat J^{(+)}(X)$ \vspace{.1 cm}\\}}}
\end{center}

In this section we are interested in obtaining analytical expressions that
permit us to compare the proposed probability distribution of the time of
arrival $\langle \psi _{{\rm tr}}(t)|\hat J^{(+)}(X)|\psi _{{\rm tr}%
}(t)\rangle $ with the standard probability current density $\langle \psi _{%
{\rm tr}}(t)|\hat J(X)|\psi _{{\rm tr}}(t)\rangle $. To this end we shall
restrict ourselves to a free particle characterized, at $t=0$, by a minimum
Gaussian wave packet with centroid $x_0$, having a negligible contribution
of negative-momentum components, and propagating with average momentum $%
p_0>0 $ along the $x$ axis toward a detector located at a certain position $%
X>x_0$. Specifically,

\begin{equation}
\label{gauspx1}\langle p|\psi (0)\rangle =\left[ 2\pi (\Delta p)^2\right]
^{-1/4}\exp \left[ -\left( \frac{p-p_0}{2\Delta p}\right) ^2-i\frac{px_0}%
\hbar \right] ,
\end{equation}
where the momentum spread $\Delta p\ll p_0$ is assumed to be sufficiently
small so as to satisfy $\langle p|\psi (0)\rangle \simeq \Theta (p)\langle
p|\psi (0)\rangle $ to a good approximation. As stated above, under these
conditions $T(p)\rightarrow 1$ and we may substitute $|\psi _{{\rm tr}%
}(t)\rangle \rightarrow |\psi (t)\rangle $ throughout the relevant formulas.
The integrals involved in the definition of the probability current density
[Eqs. (\ref{prim}) and (\ref{funt})] can then be easily carried out to
obtain the well-known formula

\begin{equation}
\label{jqrm}\langle \psi (t)|\hat J(X)|\psi (t)\rangle =\frac{\sqrt{2/\pi }%
\Delta p}{m\hbar }\frac{\left( p_0+4\frac{(\Delta p)^4(X-x_0)}{m\hbar ^2}%
t\right) }{\left( 1+4\frac{(\Delta p)^4}{m^2\hbar ^2}t^2\right) ^{3/2}}\exp
\left( -2\frac{(\Delta p)^2}{\hbar ^2}\frac{\left[ (X-x_0)-\frac{p_0}%
mt\right] ^2}{1+4\frac{(\Delta p)^4}{m^2\hbar ^2}t^2}\right) . 
\end{equation}
As far as the probability distribution $\langle \psi (t)|\hat
J^{(+)}(X)|\psi (t)\rangle $ is concerned, some additional simplification is
needed. We shall content ourselves with an analytical approximation valid up
to order $(\Delta p/p_0)^2$. To this end, following Grot {\em et al. }[\ref
{Grot}], we expand the argument of the functional $I[\sqrt{p}]$ entering Eq.
(\ref{segu}) as a Taylor series about the average momentum $p_0$, i.e.,

\begin{equation}
\label{des1}\sqrt{p}=\sqrt{p_0}\left[ 1+\frac{p-p_0}{2p_0}-\frac 12\left( 
\frac{p-p_0}{2p_0}\right) ^2+O\left( \frac{p-p_0}{p_0}\right) ^3\right] . 
\end{equation}
Substitution of this expansion into $I[\sqrt{p}]$ leads to

\begin{equation}
\label{vtta}I[\sqrt{p}]=\frac 14\sqrt{p_0}\left[ \frac 32I[1]+\frac
3{p_0}I[p]-\frac 1{2p_0^2}I[p^2]+O\left( \frac{\Delta p}{p_0}\right)
^3\right] . 
\end{equation}
The important point is that both $I[p]$ and $I[p^2]$ can be written in terms
of $I[1]$, yielding an expression for $\langle \psi (t)|\hat J^{(+)}(X)|\psi
(t)\rangle $ that can be easily related to the probability current density
given in Eq. (\ref{jqrm}). Indeed, substituting Eq. (\ref{gauspx1}) in the
integrand of $I[1]$ and taking into account that $\langle p|\psi (0)\rangle
\simeq 0$ for $p<0$, one arrives at the Gaussian integral

\begin{equation}
\label{funt1}I[1]=\!N\int_{-\infty }^{+\infty }\!\!dp\,e^{-\delta \left(
p-\lambda \right) ^2}, 
\end{equation}
where

\begin{equation}
\label{lpri1}\delta \equiv \frac 1{4(\Delta p)^2}+i\frac t{2m\hbar }, 
\end{equation}

\begin{equation}
\label{lpri2}\lambda \equiv \frac{p_0+2i(\Delta p)^2(X-x_0)/\hbar }{%
1+2i(\Delta p)^2t/m\hbar }.
\end{equation}
(The factor $N$ is not relevant to our purposes and consequently is not
given here.) Note that ${\rm Re}(\delta )>0$, as required for the integral
to converge. As is well known, the integral of Eq. (\ref{funt1})  is, in
fact, independent of $\lambda $. Thus, by differentiating $I[1]$ with
respect to $\lambda $ one can readily show that

\begin{equation}
\label{lep1}I[p]=\lambda I[1]. 
\end{equation}
Likewise, a second differentiation with respect to $\lambda $ leads to

\begin{equation}
\label{lep2}I[p^2]=\left( \lambda ^2+\frac 1{2\delta }\right) I[1].
\end{equation}
By inserting Eq. (\ref{lep1}) into Eq. (\ref{prim}), one obtains

\begin{equation}
\label{vtxs}\langle \psi (t)|\hat J(X)|\psi (t)\rangle =\frac 1{mh}{\rm Re}%
(\lambda )\,|I[1]|^2.
\end{equation}
Substitution of Eqs. (\ref{lep1}) and (\ref{lep2}) into Eq. (\ref{vtta})
yields an expression for $I[\sqrt{p}]$ depending only on $I[1]$. Inserting
then the expression so obtained into Eq. (\ref{segu}) and using Eq. (\ref
{vtxs}) to eliminate $\,|I[1]|^2$ in favor of the probability current $%
\langle \psi (t)|\hat J(X)|\psi (t)\rangle $, we arrive at

\begin{eqnarray}
\langle \psi (t)|\hat J^{(+)}(X)|\psi (t)\rangle =\frac 3{16p_0^3\,{\rm Re}
(\lambda )}\left\{ \frac 34p_0^4+3p_0^3\,{\rm Re}(\lambda )+3p_0^2\,|\lambda
|^2-\frac 12p_0^2\,{\rm Re}\left( \lambda ^2+\frac 1{2\delta }\right)
\right. \nonumber \\
\label{nsx} - \left. p_0\,{\rm Re}\left[\lambda \left(\lambda^2+
\frac 1{2\delta
}\right) ^{*}\right]+\frac 1{12}{\bf |}\lambda ^2+\frac 1{2\delta }
{\bf |}%
^2+O\left( \frac{\Delta p}{p_0}\right) ^3\right\} \langle \psi (t)|\hat
J(X)|\psi (t)\rangle .
\end{eqnarray}
Of course, for this cumbersome expression to be useful some simplification
is still required. Let $t_0=(X-x_0)m/p_0$ be the classical time of arrival
at the detector located at $X$. Restricting ourselves to particles arriving
in the time interval $[0,2t_0]$, we have

\begin{equation}
\label{ntic}t\leq 2t_0\;\;\Rightarrow \;\;\frac{(\Delta p)^2t}{m\hbar }\leq
\rho \left( \frac{\Delta p}{p_0}\right) , 
\end{equation}
where the parameter $\rho $ denotes the distance between the centroid of the
wave packet at $t=0$ and the detector's position, in units of the spatial
spread $\Delta x=\hbar /2\Delta p$, i.e., $(X-x_0)\equiv \rho \Delta x$. The
important point is that when $\rho \sim O(1)$, the term $2(\Delta
p)^2t/m\hbar $ involved in the expressions of both $1/\delta $ and $\lambda $
[Eqs. (\ref{lpri1}) and (\ref{lpri2})] becomes of order $(\Delta p/p_0)$.
This fact enables us to approximate the various terms contributing to Eq. (%
\ref{nsx}) as Taylor's expansions up to terms of order $(\Delta p/p_0)^2$.
For instance, under the assumptions just stated we would have

\begin{equation}
\label{algst}{\rm Re}(\lambda )=p_0\left[ 1+4\frac{(\Delta p)^4(X-x_0)t}{%
mp_0\hbar ^2}-4\frac{(\Delta p)^4t^2}{m^2\hbar ^2}+O\left( \frac{\Delta p}{%
p_0}\right) ^3\right] , 
\end{equation}
with similar expansions for the rest of the terms entering the above
expression for $\langle \psi (t)|\hat J^{(+)}(X)|\psi (t)\rangle $. The
substitution of all of these expansions into Eq. (\ref{nsx}) leads, after a
rather lengthy calculation, to the final result

\begin{equation}
\label{ntcl}\langle \psi (t)|\hat J^{(+)}(X)|\psi (t)\rangle =\left[
1+\Lambda _0-\Lambda _1t+\Lambda _2t^2+O\left( \frac{\Delta p}{p_0}\right)
^3\right] \langle \psi (t)|\hat J(X)|\psi (t)\rangle , 
\end{equation}
where the coefficients $\Lambda _0$, $\Lambda _1$, and $\Lambda _2$ are
defined as

\begin{equation}
\label{co1e}\Lambda _0\equiv -\frac 12\left( \frac{\Delta p}{p_0}\right) ^2+%
\frac{2(\Delta p)^4(X-x_0)^2}{\hbar ^2p_0^2}, 
\end{equation}

\begin{equation}
\label{co2e}\Lambda _1\equiv 4\frac{(\Delta p)^4(X-x_0)}{mp_0\hbar ^2}, 
\end{equation}

\begin{equation}
\label{co3e}\Lambda _2\equiv 2\frac{(\Delta p)^4}{m^2\hbar ^2}. 
\end{equation}

Equation (\ref{ntcl}) constitutes the main result of this section. When used
along with Eq. (\ref{jqrm}), it enables us to obtain analytically the
probability distribution $\langle \psi (t)|\hat J^{(+)}(X)|\psi (t)\rangle $
up to order $(\Delta p/p_0)^2$. When the detector is located at a position $X
$ initially separated from the centroid $x_0$ a distance of order $\Delta x$%
, its range of applicability extends over the time interval $[0,2t_0]$.
However, the validity of Eq. (\ref{ntcl}) is not restricted to this
particular configuration. Indeed, it is not difficult to see that for any
detector location $X>x_0$ (i.e., any $\rho >0$), Eq. (\ref{ntcl}) remains
true within the interval $[0,\sigma t_0]$, with $\sigma \approx \min \left\{
(2/\rho ),(2/\rho )^2\right\} $. For greater times the contribution from
those terms neglected in the various expansions increases, so that they can
no longer be considered to be of order $(\Delta p/p_0)^3$ and Eq. (\ref{ntcl}%
) might fail as an approximation valid up to order $(\Delta p/p_0)^2$.

As is apparent from the above formulas, the difference between the
expectation values of $\hat J(X)$ and $\hat J^{(+)}(X)$ turns out to be of
order $(\Delta p/p_0)^2$, so that, in the limit $(\Delta p/p_0)\rightarrow 0$%
, both of them coincide, a result that is in agreement with Ref. [\ref{Grot}%
] and confirms the asymptotic analysis of Ref. [\ref{VDB5}]. A comparison
between these two quantities is, however, most conveniently done in terms of
their relative difference.

\vspace{1.4 cm}

\begin{center}
{\large {\bf {IV. RELATIVE DIFFERENCE\vspace{.1 cm}\\}}}
\end{center}

Given a Hamiltonian $\hat H=\hat H_0+V(\hat X)$ and a state vector $|\psi
(0)\rangle $ having no contribution of negative-momentum components, and not
overlapping appreciably with the potential barrier [so that one may
legitimately substitute $|\psi _{{\rm in}}\rangle \rightarrow $ $|\psi
(0)\rangle $], the relative difference $\Delta $ between the probability
distribution $\langle \psi _{{\rm tr}}(t)|\hat J^{(+)}(X)|\psi _{{\rm tr}%
}(t)\rangle $ and the corresponding probability current density $\langle
\psi _{{\rm tr}}(t)|\hat J(X)|\psi _{{\rm tr}}(t)\rangle $ can be written as
[see Eq. (\ref{ptrix})]

\begin{equation}
\label{npid}\Delta =1-\frac{\langle \psi (0)|\,{\hat S}^{\dagger }\Theta
(\hat P)\hat J_{{\rm I}}(X,t)\Theta (\hat P){\hat S}\,|\psi (0)\rangle }{%
\langle \psi (0)|\,{\hat S}^{\dagger }\Theta (\hat P)\hat J_{{\rm I}%
}^{(+)}(X,t)\Theta (\hat P){\hat S}\,|\psi (0)\rangle }. 
\end{equation}
This quantity, which constitutes the basis for our subsequent analysis, is
(for any $t$) a state functional that enables us to quantify the differences
we are interested in, as a function of the initial state $|\psi (0)\rangle $%
. To this end we shall consider an initial state vector of the form

\begin{equation}
\label{suprx}|\psi (0)\rangle =\alpha \left( \beta \,|\psi _1\rangle +|\psi
_2\rangle \right) , 
\end{equation}
where $\alpha $ is the normalization constant,

\begin{equation}
\label{cche}\alpha =\left( \beta ^2+2\beta \,{\rm Re}\langle \psi _1|\psi
_2\rangle +1\right) ^{-1/2},
\end{equation}
and $\beta $ is an arbitrary real coefficient. On the other hand, for
computational simplicity we shall choose $\langle p|\psi _j\rangle $ $(j=1,2)
$ to be minimum Gaussian wave packets centered at $x_0$, with momentum
spread $\Delta p$  and average momentum $p_j$, respectively,

\begin{equation}
\label{gauspx}\langle p|\psi _j\rangle =\left[ 2\pi (\Delta p)^2\right]
^{-1/4}\exp \left[ -\left( \frac{p-p_j}{2\Delta p}\right) ^2-i\frac{px_0}%
\hbar \right] .
\end{equation}
Furthermore, we take $p_2\geq p_1>0$ and $\Delta p\ll p_1$ in order to
guarantee that $|\psi (0)\rangle $ has no appreciable contribution of
negative-momentum components.

The interest in choosing $|\psi (0)\rangle $ this way comes from the fact
that by varying the continuous parameters $\beta $, $p_1$, $p_2$, $\Delta p$%
, and $x_0$ we can easily explore different regions of the Hilbert space.
Such an analysis will be the aim of the next section. For the time being,
consider a free particle described at  $t=0$ by the state vector defined by
Eqs. (\ref{suprx})--(\ref{gauspx}) with $p_2\rightarrow p_1\equiv $ $p_0$.
In this particular case, $|\psi (0)\rangle \rightarrow |\psi _1\rangle $ and
the initial state reduces to the simple minimum Gaussian wave packet
considered in the preceding section [Eq. (\ref{gauspx1})]. Under these
circumstances, an analytical expression can be derived for $\Delta $. By
substituting Eq. (\ref{ntcl}) into Eq. (\ref{npid}), one finds (for $0\leq
t\leq \sigma t_0$)

\begin{equation}
\label{dunp}\Delta =\Lambda _2(t-t_0)^2-\frac 12\left( \frac{\Delta p}{p_0}%
\right) ^2+O\left( \frac{\Delta p}{p_0}\right) ^3, 
\end{equation}
with $\Lambda _2$ given in Eq. (\ref{co3e}). Therefore, as a function of $t$%
, the relative difference $\Delta $ is given by a parabola which reaches its
minimum at the classical arrival time $t_0=(X-x_0)m/p_0$. From Eq. (\ref
{dunp}) it follows that $\Delta (t_0)<0$, and consequently

\begin{equation}
\label{yaet}\langle \psi (t_0)|\hat J^{(+)}(X)|\psi (t_0)\rangle <\langle
\psi (t_0)|\hat J(X)|\psi (t_0)\rangle , 
\end{equation}
so that the probability of arriving at $t_0$ as predicted by the current $%
\hat J(X)$ is always greater than that predicted by the modulus of the
current $\hat J^{(+)}(X)$. More generally, it can be readily seen that for
any $t\geq 0$ within the symmetric interval $[t_0-m\Delta x/p_0,t_0+m\Delta
x/p_0]$ about $t_0$, it holds that

\begin{equation}
\label{dunim}\langle \psi (t)|\hat J^{(+)}(X)|\psi (t)\rangle \leq \langle
\psi (t)|\hat J(X)|\psi (t)\rangle , 
\end{equation}
where the equality is satisfied at the boundaries of the interval [up to
terms of order $(\Delta p/p_0)^2$].

\vspace{1.4 cm}

\begin{center}
{\large {\bf {V. QUANTITATIVE ANALYSIS\vspace{.2 cm}\\}}}
\end{center}


\begin{center}
{\large {\bf {A. Free particle\vspace{.1 cm}\\}}}
\end{center}

We begin the analysis of the relative difference $\Delta $ between the
probability distribution $\langle \psi _{{\rm tr}}(t)|\hat J^{(+)}(X)|\psi _{%
{\rm tr}}(t)\rangle $ and the probability current density $\langle \psi _{%
{\rm tr}}(t)|\hat J(X)|\psi _{{\rm tr}}(t)\rangle $ by considering a free
particle described at $t=0$ by the minimum Gaussian wave packet defined in
Eq. (\ref{gauspx1}) [which, as already said, is nothing but a particular
case of the state vector previously introduced in Eqs. (\ref{suprx})--(\ref
{gauspx})]. To be specific, we shall restrict our investigation to an
electron with average momentum $p_0=0.5$ a.u., and place the detector at $%
X=x_0+3\Delta x$, where $\Delta x=\hbar /2\Delta p$ is the spatial spread of 
$|\psi (0)\rangle $.

Figures 1(a) and 1(b) show the relative difference $\Delta $, as a function
of the time of arrival, for several values of $\Delta p$ ranging over two
orders of magnitude. The detection interval has been chosen to be symmetric
about the classical arrival time $t_0$, i.e., $t\in \left[ 0,2t_0\right] $.
Moreover, for comparison purposes the final instant of time has been
normalized to unity in all cases (i.e., $t\rightarrow t_{{\rm n}}\equiv
t/2t_0$), so that $t_0$ lies always in the middle of the detection interval
considered. It is worth remarking that the probability current density turns
out to be positive in all of the cases studied.

Besides the results obtained from a direct numerical integration of the
corresponding formulas, Fig. 1(b) also shows the behavior of $\Delta $ as
predicted by Eq. (\ref{dunp}). (Only the theoretical values corresponding to 
$\Delta p=0.01$ a.u. have been explicitly plotted [circles in Fig. 1(b)]
since for smaller  $\Delta p$ the agreement is even better.) As is apparent
from this figure, within the range of validity of the theoretical prediction
(that is, for $(\Delta p/p_0)\ll 1$ and $t\in [0,\sigma t_0]$) the agreement
turns out to be excellent (as it should be). In fact, for any momentum
uncertainty $\Delta p\leq 0.01$ a.u., the relative difference between $%
\langle \hat J^{(+)}(X)\rangle (t)$ and $\langle \hat J(X)\rangle (t)$,
considered as a function of $t$, behaves as a parabola which cuts the
horizontal axis at $t_{{\rm n}}=1/3$ and $t_{{\rm n}}=2/3$, reaching its
minimum value at the classical time $t_{{\rm n}}=1/2$. Furthermore, while
within the interval $[1/3,2/3]$ the expectation value of $\hat J(X)$ always
dominates over the expectation value of $\hat J^{(+)}(X)$, just the opposite
occurs outside such an interval.

From Figs. 1(a) and 1(b) the rapid decrease of $\Delta $ with the momentum
spread is also apparent,  which is a direct consequence of the fact that for 
$\Delta p/p_0$ sufficiently small, $\Delta $ is of order $(\Delta p/p_0)^2$.
In this regard, note that for $\Delta p\leq 0.01$ a.u. the relative
difference is already less than $0.2\%$ over all of the  detection interval
considered, and that figure would be even smaller if one focused attention
on a time interval more localized about the most probable time of arrival $%
t_0$. This fact would render any attempt to discriminate between $\langle
\hat J^{(+)}(X)\rangle (t)$ and $\langle \hat J(X)\rangle (t)$ almost
impossible in practice. Such a negative conclusion should not be
extrapolated, however. A minimum Gaussian wave packet represents a very
special type of quantum state, for it exhibits the lowest possible
uncertainty product $\Delta x\Delta p$ and consequently is expected to be
largely semiclassical in character. On the other hand, as can be inferred
from an asymptotic analysis [\ref{VDB5}], in the semiclassical limit $\hbar
\rightarrow 0$ it holds that

\begin{equation}
\label{dalv}\langle \psi (t)|\hat J^{(+)}(X)|\psi (t)\rangle \rightarrow
|\langle \psi (t)|\hat J(X)|\psi (t)\rangle |,
\end{equation}
so that the small value found for $\Delta $ in the cases considered above is
not surprising.

A simple way for generating initial states having a more genuine quantum
character consists in allowing $p_1$ to be different from $p_2$ in Eqs. (\ref
{suprx})--(\ref{gauspx}). In fact, a marked interference pattern can be
induced in the probability current density (as well as in the corresponding
position probability density) by simply increasing the distance (in momentum
space) between $p_1$ and $p_2$ while keeping unchanged the remaining
parameters.

Figures 2 and 3 show the results obtained by taking $\beta =2$, $p_2=0.5$
a.u., $\Delta p=0.01$ a.u., and allowing $p_1$ to vary between $0.4$ and $0.2
$ a.u., respectively. The detector's position has been chosen as before
(more precisely, we have taken $X-x_0=3\hbar /2\Delta p$) and the detection
interval (now expressed in atomic units) has been chosen in such a way that
its final time $t_{{\rm f}}$ satisfies

\begin{equation}
\label{otem}\langle \psi (t_{{\rm f}})|\hat J^{(+)}(X)|\psi (t_{{\rm f}%
})\rangle \approx \langle \psi (0)|\hat J^{(+)}(X)|\psi (0)\rangle . 
\end{equation}

In Figs. 2(b) and 3(b) we have plotted the expectation values of both $\hat
J^{(+)}(X)$ and $\hat J(X)$ corresponding, respectively, to $p_1=0.4$ a.u.
and $p_1=$ $0.2$ a.u. These curves can be considered as being obtained by
means of a continuous deformation (induced by varying $p_1$) starting from
the initial Gaussian profile corresponding to $p_1=p_2=0.5$ a.u. Since a
marked interference pattern is a hallmark of quantum behavior, it is evident
that by decreasing $p_1$ we are probing domains of the Hilbert space with
increasing quantum character. As is apparent from these figures, while for $%
p_1=0.4$ a.u. the probability current density remains positive over all of
the time interval considered, the same does not occur for $p_1=0.2$ a.u.,
and in this case $\langle \psi (t)|\hat J(X)|\psi (t)\rangle $ attains
negative values in the neighborhood of $t=40$ a.u. and $t=400$ a.u.

The analysis of the corresponding relative differences, plotted in Figs.
2(a) and 3(a), reveals that  now $\Delta $ exhibits a series of maxima
(clearly related to the interference pattern of the probability current
density) where it can take values of order $10\%$ (in the first case) or $%
100\%$ (in the second one). [In fact, in this latter case the maxima which
have been truncated in the figure correspond, respectively, to $\Delta (39.1$
a.u.$)=31.96$ and $\Delta (398.05$ a.u.$)=1011.58$.] Therefore, the
probability distributions of the time of arrival as predicted by $\langle
\hat J^{(+)}(X)\rangle (t)$ or $\langle \hat J(X)\rangle (t)$ can be quite
different in these cases. A comparison with the bound $0.2\%$ obtained
previously for $p_1=p_2=0.5$ a.u. [Fig. 1(b)] reflects the fact that the
choice of the initial state plays an essential role in both the magnitude
and the behavior of the relative discrepancy between $\langle \hat
J^{(+)}(X)\rangle (t)$ and $\langle \hat J(X)\rangle (t)$.

An interesting limiting situation, where quantum effects dominate the
behavior of the probability current density, can be achieved by taking $%
\Delta p\rightarrow 0$ and $p_2/p_1\gg \beta \gg 1$ in Eqs. (\ref{suprx})--(%
\ref{gauspx}). In this particular case, the initial state becomes a purely
quantum state with no classical analog, consisting of the coherent
superposition of two macroscopically distinguishable states in momentum
space. Under these circumstances, the main contribution to $\langle \psi
(t)|\hat J(X)|\psi (t)\rangle $ comes from the interference terms and one
obtains, to leading order as $\Delta p\rightarrow 0$ [\ref{VDB5}],

\begin{equation}
\label{doft}\langle \psi (t)|\hat J(X)|\psi (t)\rangle \sim \frac{2\sqrt{%
2\pi }}{mh}\beta \alpha ^2\Delta p\,\,p_2\cos \left[ p_2^2t/2\hbar
m-p_2(X-x_0)/\hbar \right] +O\left[ (\Delta p)^2\right] .
\end{equation}
This case is illustrated in Figs. 4(a) and 4(b), where we have specifically
taken $\beta =100$, $p_2=1$ a.u., $p_1=4\times 10^{-3}$ a.u., $\Delta
p=5\times 10^{-4}$ a.u., and $X-x_0=3\hbar /2\Delta p$. Also plotted in Fig.
4(b) is the modulus of the probability current density, $|\langle \psi
(t)|\hat J(X)|\psi (t)\rangle |$, which is the proposal by McKinnon and
Leavens [\ref{Leav3}] for the probability distribution of the time of
arrival.

From a comparison between Figs. 4(a) and 4(b) it can be seen that the
relative difference $\Delta $ between $\langle \hat J(X)\rangle (t)$ and $%
\langle \hat J^{(+)}(X)\rangle (t)$ reaches a local maximum exactly when $%
\langle \hat J(X)\rangle (t)$ reaches a local minimum, and in this case $%
\Delta \gg 1$, so that

\begin{equation}
\label{ahms}\Delta \approx -\frac{\langle \hat J(X)\rangle (t)}{\langle \hat
J^{(+)}(X)\rangle (t)}=\frac{{\bf |}\langle \hat J(X)\rangle (t)|}{\langle
\hat J^{(+)}(X)\rangle (t)}.
\end{equation}
Consequently, the main effect produced by the replacement $\langle \hat
J(X)\rangle (t)\rightarrow |\langle \hat J(X)\rangle (t)|$ upon the
corresponding relative difference would consist in a change of sign. A
similar conclusion could be inferred from Fig. 3(a) for those instants of
time where the probability current density becomes negative. The important
consequence is that, from a quantitative setting, by replacing the
probability current density by its modulus one in general does not achieve a
better agreement with the probability distribution $\langle \hat
J^{(+)}(X)\rangle (t)$.

\vspace{1.2 cm}

\begin{center}
{\large {\bf {B. Potential barrier\vspace{.1 cm}\\}}}
\end{center}

The presence of a potential barrier might, in principle, induce some
additional discrepancy between $\langle \hat J(X)\rangle (t)$ and $\langle
\hat J^{(+)}(X)\rangle (t)$. In order to examine whether this is the case,
we shall consider next a quantum particle propagating toward a detector
located at a certain asymptotic point $X$ behind a one-dimensional potential
barrier. It turns out to be most convenient restricting to initial states as
simple as possible since in these cases the effect produced specifically by
the barrier can be more easily identified. Accordingly, we shall consider an
electron characterized at $t=0$ by the Gaussian state given in Eq. (\ref
{gauspx1}), with $p_0=0.5$ a.u. The potential barrier is assumed to occupy
the segment $[0,d]$ of the $x$ axis, and its height $V_0$ has been chosen in
all cases so that $p_B\equiv \sqrt{2mV_0}=0.8$ a.u.

Our primary interest consists in investigating the behavior of the relative
difference $\Delta $ as a function of both the barrier's width $d$ and the
momentum spread $\Delta p$. To this end we have selected the centroid $x_0$
of the initial wave packets in such a way that $\langle x|\psi (0)\rangle $
does not overlap appreciably (in comparison with the transmittance ${\sf T}$%
) with the interaction center. Specifically, $x_0$ has been implicitly
defined by

\begin{equation}
\label{ducg}\int_0^\infty dx|\langle x|\psi (0)\rangle |^2\simeq 10^{-3}{\sf %
T}
\end{equation}
and, consequently, is a function of both $\Delta p$ and $d$. On the other
hand, in order to guarantee the applicability of the formalism (which
requires the particle to be asymptotically free), the detector has  been
assumed to be switched on at a certain instant $t_{{\rm i}}$ satisfying the
condition that the probability of finding the particle within the
interaction region is already negligible. More precisely, we have defined
implicitly $t_{{\rm i}}$ by the condition

\begin{equation}
\label{stdq}\int_{-\infty }^ddx|\langle x|\psi _{{\rm tr}}(t_{{\rm i}%
})\rangle |^2\simeq 10^{-3}{\sf T.}
\end{equation}
The detector has also been assumed to be located at a certain position $X$
behind the interaction center sufficiently far from the barrier's edge so as
to satisfy the condition that when it is switched on (at $t=t_{{\rm i}}$)
the probability of finding (in the absence of detector) the transmitted
particle within the region $x\geq X$ is still negligible. That is, $X$ has
been obtained from the condition

\begin{equation}
\label{panh}\int_X^\infty dx|\langle x|\psi _{{\rm tr}}(t_{{\rm i}})\rangle
|^2\simeq 10^{-3}{\sf T.} 
\end{equation}
Finally, it has been assumed that the detector is switched off at a certain
instant $t_{{\rm f}}$ satisfying

\begin{equation}
\label{doub}\langle \psi _{{\rm tr}}(t_{{\rm f}})|\hat J^{(+)}(X)|\psi _{%
{\rm tr}}(t_{{\rm f}})\rangle \approx \langle \psi _{{\rm tr}}(t_{{\rm i}%
})|\hat J^{(+)}(X)|\psi _{{\rm tr}}(t_{{\rm i}})\rangle . 
\end{equation}

Combining Eqs. (\ref{stdq}) and (\ref{panh}) we see that at $t=t_{{\rm i}}$,
i.e., when the detector is switched on, the transmitted particle can be
found in the region between the barrier's edge and the detector's position
with a probability of $99.8\%$, a fact that guarantees the applicability of
the formulation developed in the preceding sections.


\begin{table}
\centering
\caption{Parameters corresponding to $\Delta p = 0.01$ a.u. All values in
         atomic units.} \vspace{0.2 cm} 	 
\begin{tabular}{ccccc} \hline \hline
   $d$     &   $x_0$    &  $t_{\rm i}$  &  $t_{\rm f}$   & $X$    \\
\hline
    2    &   $-$201.8   &    785   &   1550   &   379.0    \\
    4    &   $-$228.0   &    839   &   1600   &   382.0    \\
    8    &   $-$275.4   &    933   &   1700   &   386.5    \\
   12    &   $-$316.6   &   1014   &   1800   &   391.0    \\           
\hline \hline
\end{tabular}
\label{tabla1}
\end{table}

Figure 5 shows the relative difference $\Delta $ obtained by taking $\Delta
p=0.01$ a.u. and allowing the barrier width to vary between $d=2$ a.u. and $%
d=12$ a.u. The corresponding values for the parameters $x_0$, $t_{{\rm i}}$, 
$t_{{\rm f}}$, and $X$ [obtained numerically from Eqs. (\ref{ducg})--(\ref
{doub})] are given in Table \ref{tabla1}. For comparison purposes, the time
interval $[t_{{\rm i}},t_{{\rm f}}]$ has been renormalized to the interval $%
[0,1]$ by means of the mapping $t\rightarrow t_{{\rm n}}\equiv (t-t_{{\rm i}%
})/(t_{{\rm f}}-t_{{\rm i}})$. As is apparent from Fig. 5, the potential
barrier seems to produce no special effect on $\Delta $. In fact, even
though the transmitted state $|\psi _{{\rm tr}}(t)\rangle $ depends to a
great extent on the barrier width $d$, the relative difference between the
corresponding expectation values of $\hat J(X)$ and $\hat J^{(+)}(X)$
exhibits no appreciable dependence on this parameter. Furthermore, $\Delta $
remains very small (of order $0.25\%$) in all cases.

Things are quite different, however, when the momentum spread increases. For 
$\Delta p=0.1$ a.u. (Figs. 6 and 7 and Table \ref{tabla2}), the relative
difference not only takes greater values, but also exhibits a clear
dependence on the barrier width. From Fig. 6(a) we see that for $d=8$ a.u., $%
\Delta $ manifests an oscillatory behavior which [from a comparison with
Fig. 6(b)] can be straightforwardly related to the emergence of an incipient
interference pattern in the corresponding probability current density. The
appearance of such an interference pattern can in turn be traced back to the
interference dynamically induced between tunneling and over-the-barrier
contributions in the corresponding transmitted wave packet [\ref{VDB2}].

A similar behavior associated, however, with greater relative differences
can be appreciated in Figs. 7(a) and 7(b), which show the results
corresponding to a barrier width $d=10$ a.u. This particular case is
interesting for still another reason. As is shown in the inset of Fig. 7(b),
the probability current density takes in this case negative values in the
neighborhood of the arrival time $t=373$ a.u. (though, admittedly, very
small ones). This fact demonstrates that even for initial Gaussian wave
packets having no appreciable contribution of negative-momentum components,
the probability current density can take negative values at an asymptotic
point $X$ behind a one-dimensional potential barrier. This sole reason is
sufficient to invalidate (even under such special circumstances) its
interpretation as a probability distribution of arrival times. Of course,
there still exists the possibility of considering instead the modulus of the
probability current density. Were we to plot the relative difference between 
$\langle \hat J^{(+)}(X)\rangle (t)$ and ${\bf |}\langle \hat J(X)\rangle
(t)|$, the only relevant change in Fig. 7(a) would be the transformation of
the first maximum $\Delta (t=373.3$ a.u.$)=3.92$ into a minimum ($%
3.92\rightarrow -1.92$). Consequently, a considerable discrepancy would
still survive.


\begin{table}
\centering
\caption{Parameters corresponding to $\Delta p = 0.1$ a.u. All values in
         atomic units.} \vspace{0.2 cm} 	 
\begin{tabular}{ccccc} \hline \hline
   $d$     &   $x_0$    &  $t_{\rm i}$  &  $t_{\rm f}$   & $X$    \\
\hline
    2    &   $-$20.15   &    145.7   &   530.0   &   112.9    \\
    4    &   $-$22.48   &    137.0   &   470.0   &   108.5    \\
    8    &   $-$25.84   &    141.1   &   390.0   &   117.3    \\
   10    &   $-$28.30   &    254.0   &   573.0   &   229.4    \\           
\hline \hline
\end{tabular}
\label{tabla2}
\end{table}

\vspace{1.4 cm}

\begin{center}
{\large {\bf {VI. CONCLUSION\vspace{.1 cm}\\}}}
\end{center}

In the present work we have compared the proposal made in previous papers  [%
\ref{VDB4},\ref{VDB5}] for a quantum probability distribution of the time of
arrival at a certain point with that based on the probability current
density (or alternatively on its modulus), with the aim of establishing
conditions under which the proposals might be tested by experiment. To this
end, we began by obtaining (under certain particular conditions) an
analytical approximation for the expectation value of $\hat J^{(+)}(X)$
valid up to order $(\Delta p/p_0)^2$, and we have performed a quantitative
analysis of the corresponding relative differences as a function of the
initial state of the particle (both in the case of free evolution and in the
presence of an intermediate potential barrier).

We have found that quantum regime conditions produce the biggest differences
between the formulations which are otherwise near indistinguishable. In
fact, in the semiclassical regime, for electrons in quantum states with a
well-defined momentum $p_0\gg \Delta p$ (thus having a clear classical
analog), the relative discrepancy can be typically of order $0.2\%$, and
this figure would be even smaller if one restricted attention to arrival
times having an appreciable probability. Therefore, in this regime the
probability distribution proposed in Refs. [\ref{VDB4},\ref{VDB5}] becomes
indistinguishable in practice from the corresponding probability current
density. Important discrepancies only occur in the purely quantum regime,
for states of genuinely quantum character, having no classical analog.
Indeed, the appearance of important relative differences can be
straightforwardly related to the existence or emergence of a marked
interference pattern in the probability current density (or alternatively in
the position probability density), irrespective of the fact that such
quantum interference was already present in the initial state or was
generated dynamically through the temporal evolution of the system (as can
be the case in the presence of a scattering potential). Furthermore, a
closer analysis reveals that such quantum effects are mainly localized about
arrival times having a negligible probability and/or occur over short time
scales in comparison with the relevant time interval. These results indicate
that in order to discriminate conclusively among the different alternatives,
the corresponding experimental test should be performed in the quantum
regime and with sufficiently high resolution as to resolve small quantum
effects. Hopefully, the recent advances in the development of experimental
techniques as well as in the preparation and manipulation of atomic systems
will make such experiments feasible.

\vspace{1.2 cm}

\begin{center}
{\large {\bf {ACKNOWLEDGMENTS \vspace{.1 cm}\\}}}
\end{center}

This work has been supported by Project No. PI 2/95 from the Gobierno
Aut\'onomo de Canarias and Project No. PB97-1479-C02-01 from the Ministerio
de Educaci\'on y Cultura.

\vspace{1.4 cm}

\begin{center}
{\large {\bf {APPENDIX A\vspace{.1 cm}\\}}}
\end{center}

Let us introduce the state $|p-\rangle \equiv \hat \Omega _{-}|p\rangle $,
which is the solution of the Lippmann-Schwinger equation corresponding to an
outgoing plane wave $|p\rangle $, i.e.,

\begin{equation}
\label{ap1}|p-\rangle =|p\rangle +\left( p^2/2m-i0-\hat H\right) ^{-1}V(\hat
X)|p\rangle . 
\end{equation}
In terms of these states one can define the projector

\begin{equation}
\label{ap2}{\hat {{\cal P}}}\equiv \hat \Omega _{-}\Theta (\hat P)\hat
\Omega _{-}^{\dagger }=\!\int_0^\infty dp\,|p-\rangle \langle p-|, 
\end{equation}
which selects that part of a given state vector that will be finally
transmitted. Taking into account that $\hat \Omega _{\pm }^{\dagger }\hat
\Omega _{\pm }={\bf \hat 1}$ and ${\hat S=}\hat \Omega _{-}^{\dagger }\hat
\Omega _{+}$, it follows from Eq. (\ref{ap2}) that

\begin{equation}
\label{ap3}\hat \Omega _{-}^{\dagger }{\hat {{\cal P}}\,}\hat \Omega
_{+}=\Theta (\hat P){\hat S.} 
\end{equation}
By using the intertwining relations for the M\o ller operators [\ref{Taylor}%
],

\begin{equation}
\label{ap4}\hat \Omega _{\pm }^{\dagger }\,\hat H\,\hat \Omega _{\pm }=\hat
H_0, 
\end{equation}
as well as Eq. (\ref{ap3}), it can be readily shown that

\begin{equation}
\label{ap5}e^{-i\hat H_0t/\hbar }\,\Theta (\hat P){\hat S=\,}\hat \Omega
_{-}^{\dagger }{\hat {{\cal P}}\,}e^{-i\hat Ht/\hbar }\,\hat \Omega _{+}.
\end{equation}
Taking advantage of this relationship and recalling that $|\psi _{{\rm tr}%
}\rangle =\Theta (\hat P){\hat S}\,|\psi _{{\rm in}}\rangle $, we can write

\begin{equation}
\label{ap6}|\psi _{{\rm tr}}(t)\rangle =e^{-i\hat H_0t/\hbar }\,\Theta (\hat
P){\hat S}\,|\psi _{{\rm in}}\rangle =\hat \Omega _{-}^{\dagger }{\hat {%
{\cal P}}\,}|\psi (t)\rangle , 
\end{equation}
where we have used that $e^{-i\hat Ht/\hbar }\,\hat \Omega _{+}|\psi _{{\rm %
in}}\rangle =e^{-i\hat Ht/\hbar }|\psi (0)\rangle =|\psi (t)\rangle $. In
particular, for $t=0$, Eq. (\ref{ap6}) reduces to

\begin{equation}
\label{ap7}|\psi _{{\rm tr}}\rangle =\hat \Omega _{-}^{\dagger }{\hat {{\cal %
P}}\,}|\psi (0)\rangle . 
\end{equation}
Substitution of Eq. (\ref{ap6}) [or alternatively Eq. (\ref{ap7})] into Eq. (%
\ref{pqua}) leads to Eq. (\ref{dxnt}).

\newpage
\vspace{1.2 cm}

\begin{center}
{\large {\bf REFERENCES \vspace{.4cm}}}
\end{center}

\begin{enumerate}
\item  \label{Pauli} W. Pauli, in {\it Encyclopaedia of Physics}, edited by
S. Flugge (Springer, Berlin, 1958), Vol. 5/1, p. 60.

\item  \label{Ahaboh} Y. Aharonov and D. Bohm, Phys. Rev. {\bf 122}, 1649
(1961).

\item  \label{Allco} G. R. Allcock, Ann. Phys. (N.Y.) {\bf 53}, 253 (1969); 
{\bf 53}, 286 (1969); {\bf 53}, 311 (1969).

\item  \label{Kij} J. Kijowski, Rep. Math. Phys. {\bf 6}, 361 (1974).

\item  \label{Mielnik} B. Mielnik, Found. Phys. {\bf 24}, 1113 (1994).

\item  \label{Busch} P. Busch, M. Grabowski, and P. J. Lahti, Phys. Lett. A 
{\bf 191}, 357 (1994).

\item  \label{Du} R. S. Dumont and T. L. Marchioro II, Phys. Rev. A {\bf 47}%
, 85 (1993).

\item  \label{Leav3} W. R. McKinnon and C. R. Leavens, Phys. Rev. A {\bf 51}%
, 2748 (1995); C. R. Leavens, Phys. Lett. A {\bf 178}, 27 (1993).

\item  \label{Mug} J. G. Muga, S. Brouard, and D. Mac\'\i as, Ann. Phys.
(N.Y.) {\bf 240}, 351 (1995).

\item  \label{Grot} N. Grot, C. Rovelli, and R. S. Tate, Phys. Rev. A {\bf 54%
}, 4676 (1996).

\item  \label{leon} J. Le\'on, J. Phys. A {\bf 30}, 4791 (1997).

\item  \label{VDB4} V. Delgado and J. G. Muga, Phys. Rev. A {\bf 56}, 3425
(1997).

\item  \label{VDB5} V. Delgado, Phys. Rev. A {\bf 57}, 762 (1998).

\item  \label{Gian} R. Giannitrapani, Int. J. Theor. Phys. {\bf 36}, 1575
(1997).

\item  \label{NEW3} C. R. Leavens, Phys. Rev. A {\bf 58}, 840 (1998).

\item  \label{JPC} J. G. Muga, R. Sala, and J. P. Palao, Superlattices
Microstruct. {\bf 23}, 833 (1998).

\item  \label{NEW1} Y. Aharonov, J. Oppenheim, S. Popescu, B. Reznik, and W.
G. Unruh, Phys. Rev. A {\bf 57}, 4130 (1998).

\item  \label{Hall} J. J. Halliwell and E. Zafiris, Phys. Rev. D {\bf 57},
3351 (1998).

\item \label{frobi} F. Robicheaux, Phys. Rev. A {\bf 56}, 4296 (1998); 
{\bf 56}, 4032 (1998), and references therein.

\item  \label{onofrio} L. Viola and R. Onofrio, Phys. Rev. D {\bf 55}, 455
(1997); R. Onofrio and L. Viola, Mod. Phys. Lett. A {\bf 12}, 1411 (1997).

\item  \label{Lan} M. B\"uttiker and R. Landauer, Phys. Rev. Lett. {\bf 49},
1739 (1982).

\item  \label{Rev} For recent reviews on the subject, see (a) E. H. Hauge
and J. A. St\/ovneng, Rev. Mod. Phys. {\bf 61}, 917 (1989); (b) M.
B\"uttiker, in {\it Electronic Properties of Multilayers and Low-Dimensional
Semiconductor Structures}, edited by J. M. Chamberlain {\it et al.} (Plenum,
New York, 1990), p. 297; (c) R. Landauer, Ber. Bunsenges. Phys. Chem. {\bf 95%
}, 404 (1991); (d) C. R. Leavens and G. C. Aers, in {\it Scanning Tunneling
Microscopy III}, edited by R. Wiesendanger and H. J. G\"utherodt (Springer,
Berlin, 1993), pp. 105--140; (e) R. Landauer and T. Martin, Rev. Mod. Phys. 
{\bf 66}, 217 (1994).

\item  \label{Wigner2} M. Hillary, R. F. O'Connell, M. O. Scully, and E. P.
Wigner, Phys. Rep. {\bf 106}, 121 (1984).

\item  \label{Cohen} L. Cohen, J. Math. Phys. {\bf 7}, 781 (1966).

\item  \label{QF} See, for example, C. Itzykson and J. B. Zuber, {\it %
Quantum Field Theory} (McGraw-Hill, New York, 1985).

\item  \label{Raz} M. Razavy, Am. J. Phys. {\bf 35}, 955 (1967); Nuovo
Cimento B {\bf 63}, 271 (1969).

\item  \label{Ko} D. H. Kobe, Am. J. Phys. {\bf 61}, 1031 (1993).

\item  \label{Nota} In a recent paper [\ref{NEW3}] Leavens has compared the
prediction for the probability distribution of the time of arrival of a free
particle based on Bohm's theory with that based on the approach by Grot {\em %
et al. }[\ref{Grot}]. However, the particular cases considered by Leavens
are beyond the range of applicability of the approach we are interested in
here (Refs. [\ref{VDB4}] and [\ref{VDB5}]) and, consequently, the
corresponding conclusions cannot be extrapolated.

\item  \label{Taylor} J. R. Taylor, {\it Scattering Theory: The Quantum
Theory on Nonrelativistic Collisions} (Wiley, New York, 1975).

\item  \label{VDB2} V. Delgado and J. G. Muga, Ann. Phys. (N.Y.) {\bf 248},
122 (1996).
\end{enumerate}

\clearpage
\vspace{1.2 cm}

\begin{center}
{\large {\bf Figure captions \vspace{.4cm}}}
\end{center}

Fig. 1. Relative difference $\Delta $ as a function of the (normalized) time
of arrival $t_{{\rm n}}\equiv t/2t_0$ [where $t_0$, which depends on $\Delta
p$, is the classical arrival time: $t_0\equiv (X-x_0)m/p_0=3m\hbar
/2p_0\Delta p$]. The circles superimposed in the case $\Delta p=0.01$ a.u.
correspond to the theoretical prediction of Eq. (\ref{dunp}).

\vspace{0.3 cm}

Fig. 2. (a) Relative difference $\Delta $ as a function of the time of
arrival $t$. (b) Expectation values (indistinguishable on the present scale)
of $\hat J^{(+)}(X)$ and $\hat J(X)$ (in a.u.) as a function of $t$.

\vspace{0.3 cm}

Fig. 3. (a) Relative difference $\Delta $ as a function of the time of
arrival $t$. (b) Expectation values of $\hat J^{(+)}(X)$ and $\hat J(X)$ (in
a.u.) as a function of $t$.

\vspace{0.3 cm}

Fig. 4. (a) Relative difference $\Delta $ between $\langle \hat
J^{(+)}(X)\rangle (t)$ and $\langle \hat J(X)\rangle (t)$ as a function of
the time of arrival $t$. (b) $\langle \hat J^{(+)}(X)\rangle (t)$, $\langle
\hat J(X)\rangle (t)$, and $|\langle \hat J(X)\rangle (t)|$ (in $10^{-6}$
a.u.) as a function of $t$.

\vspace{0.3 cm}

Fig. 5. Relative difference $\Delta $ as a function of the (normalized) time
of arrival $t_{{\rm n}}\equiv (t-t_{{\rm i}})/(t_{{\rm f}}-t_{{\rm i}})$,
for $\Delta p=0.01$ a.u. and barrier widths $d=2$, $4$, $8$, and $12$ a.u.

\vspace{0.3 cm}

Fig. 6. (a) Relative difference $\Delta $ as a function of the (normalized)
time of arrival $t_{{\rm n}}\equiv (t-t_{{\rm i}})/(t_{{\rm f}}-t_{{\rm i}})$%
, for $\Delta p=0.1$ a.u. and $d=2$, $4$, and $8$ a.u. (b) Expectation
values (indistinguishable on the present scale) of $\hat J^{(+)}(X)$ and $%
\hat J(X)$ (in $10^{-6}$ a.u.) as a function of $t_{{\rm n}}$, for $\Delta
p=0.1$ a.u. and $d=8$ a.u.

\vspace{0.3 cm}

Fig. 7. (a) Relative difference $\Delta $ as a function of the time of
arrival $t$, for $\Delta p=0.1$ a.u. and $d=10$ a.u. (b) Expectation values
(indistinguishable on the present scale) of $\hat J^{(+)}(X)$ and $\hat J(X)$
(in $10^{-6}$ a.u.) as a function of $t$, for $\Delta p=0.1$ a.u. and $d=10$
a.u.

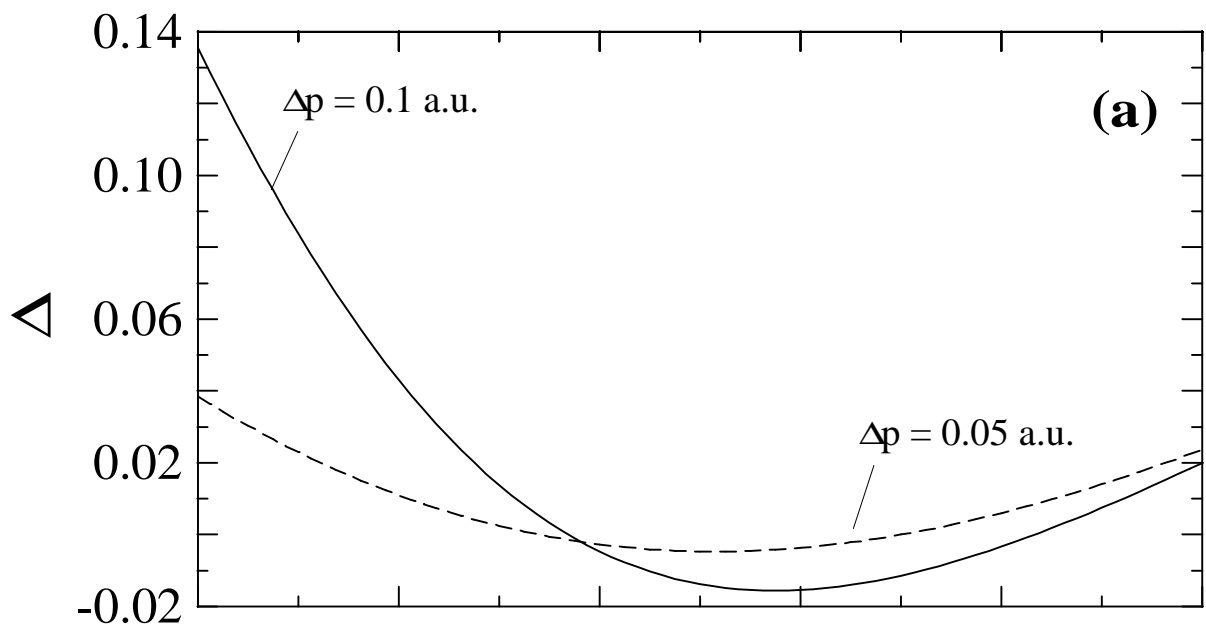
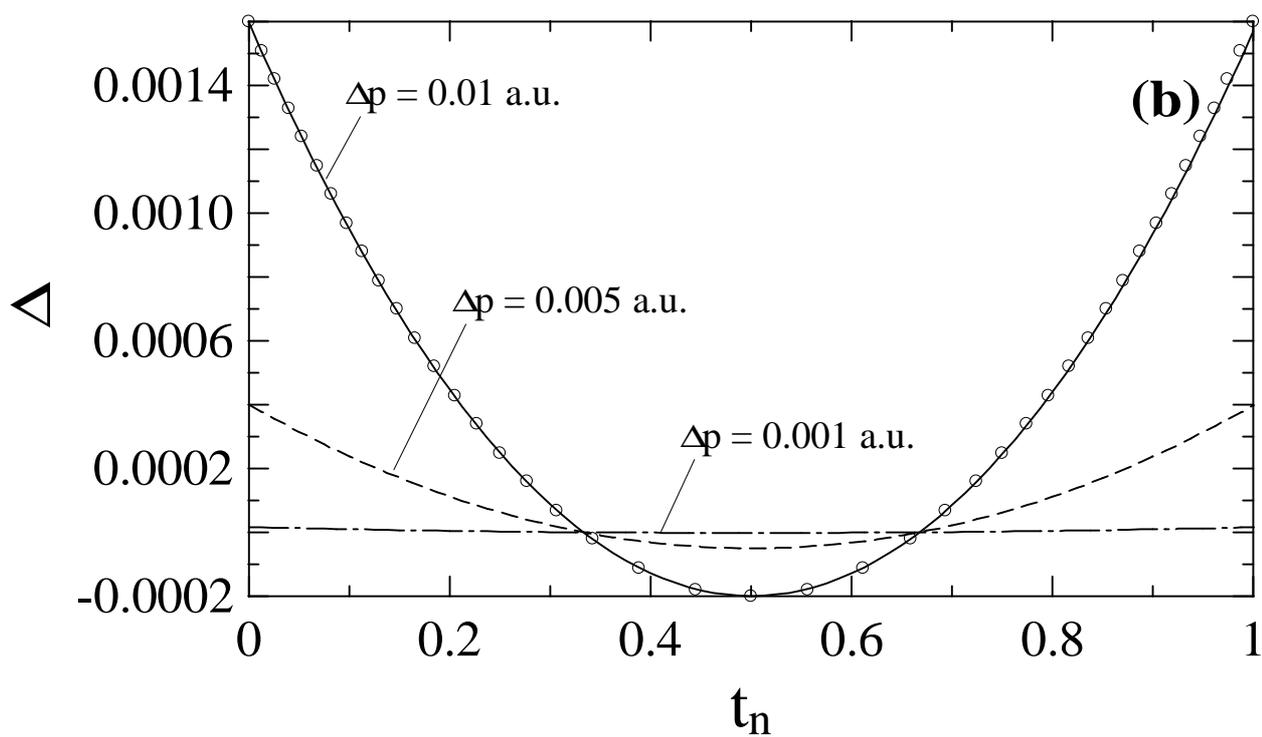
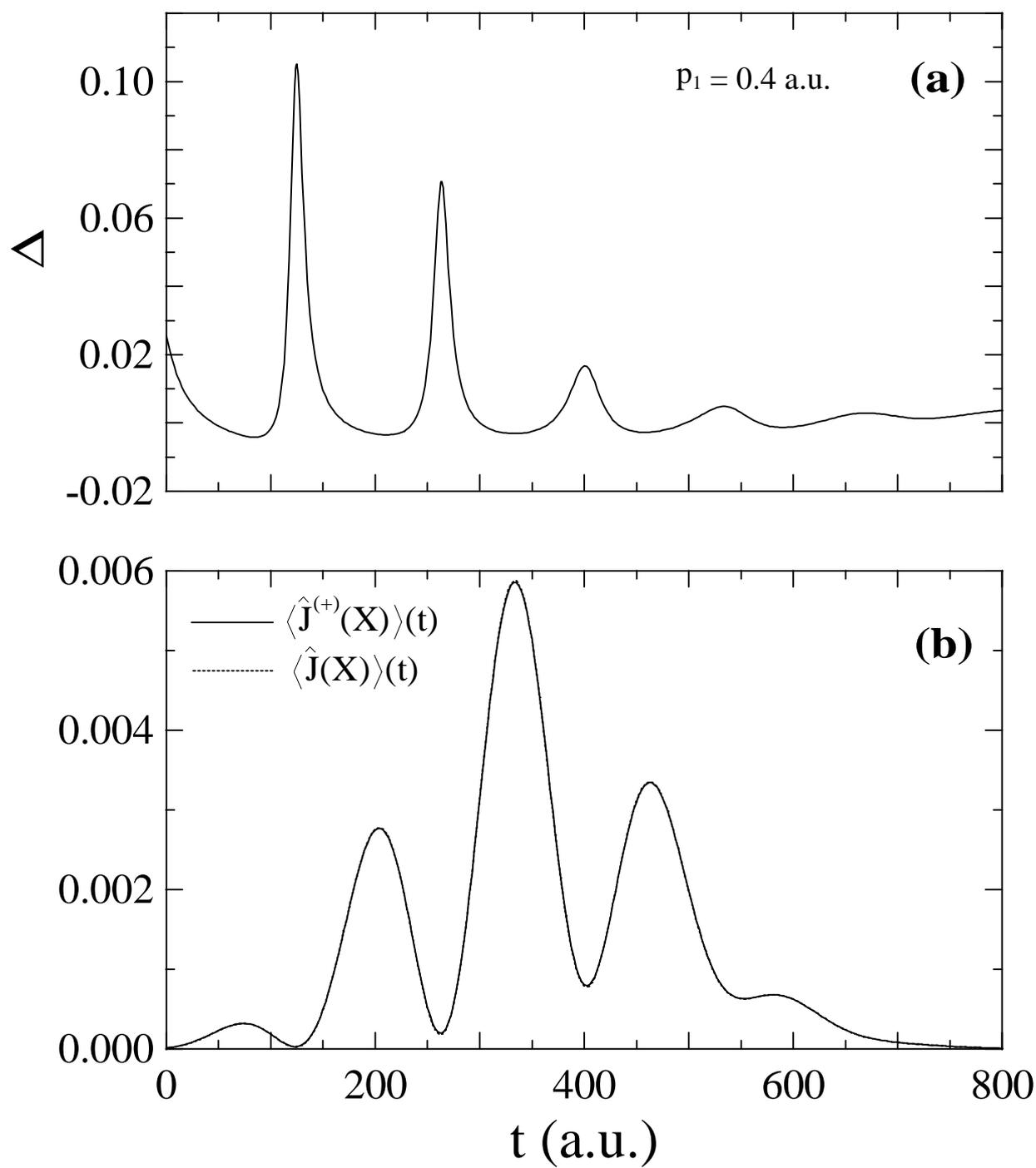
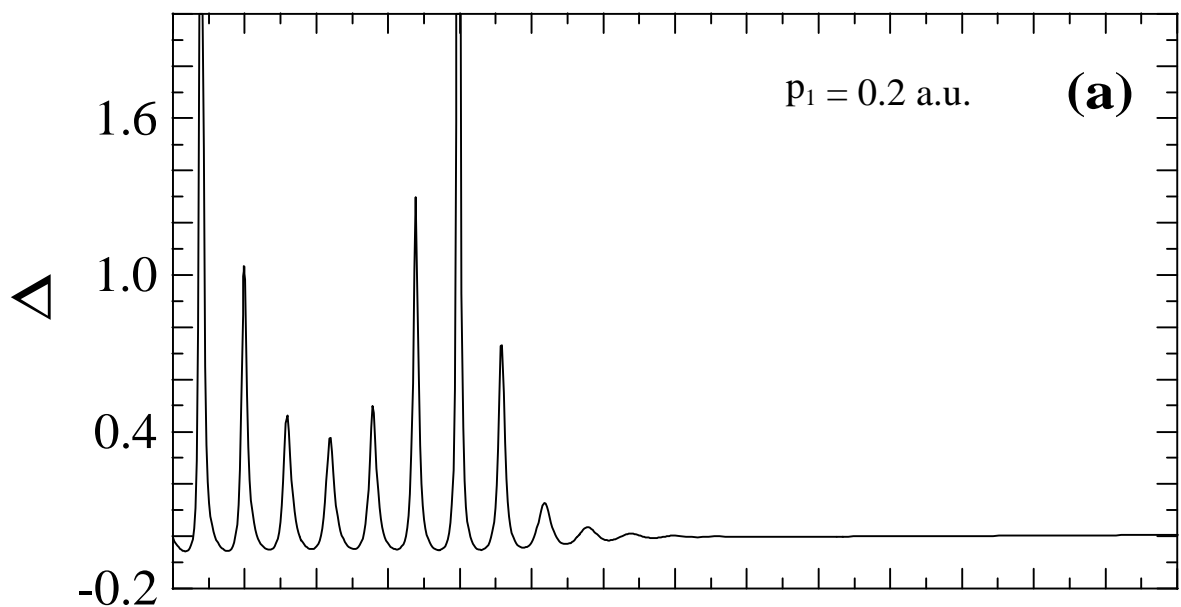
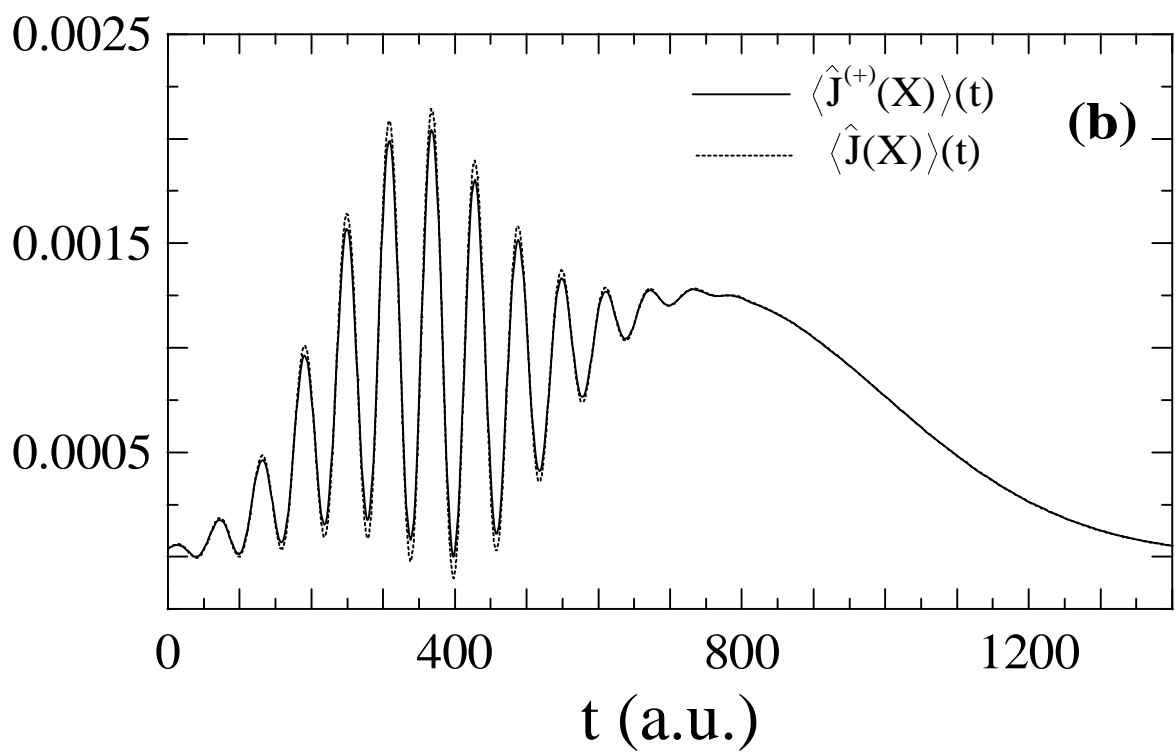
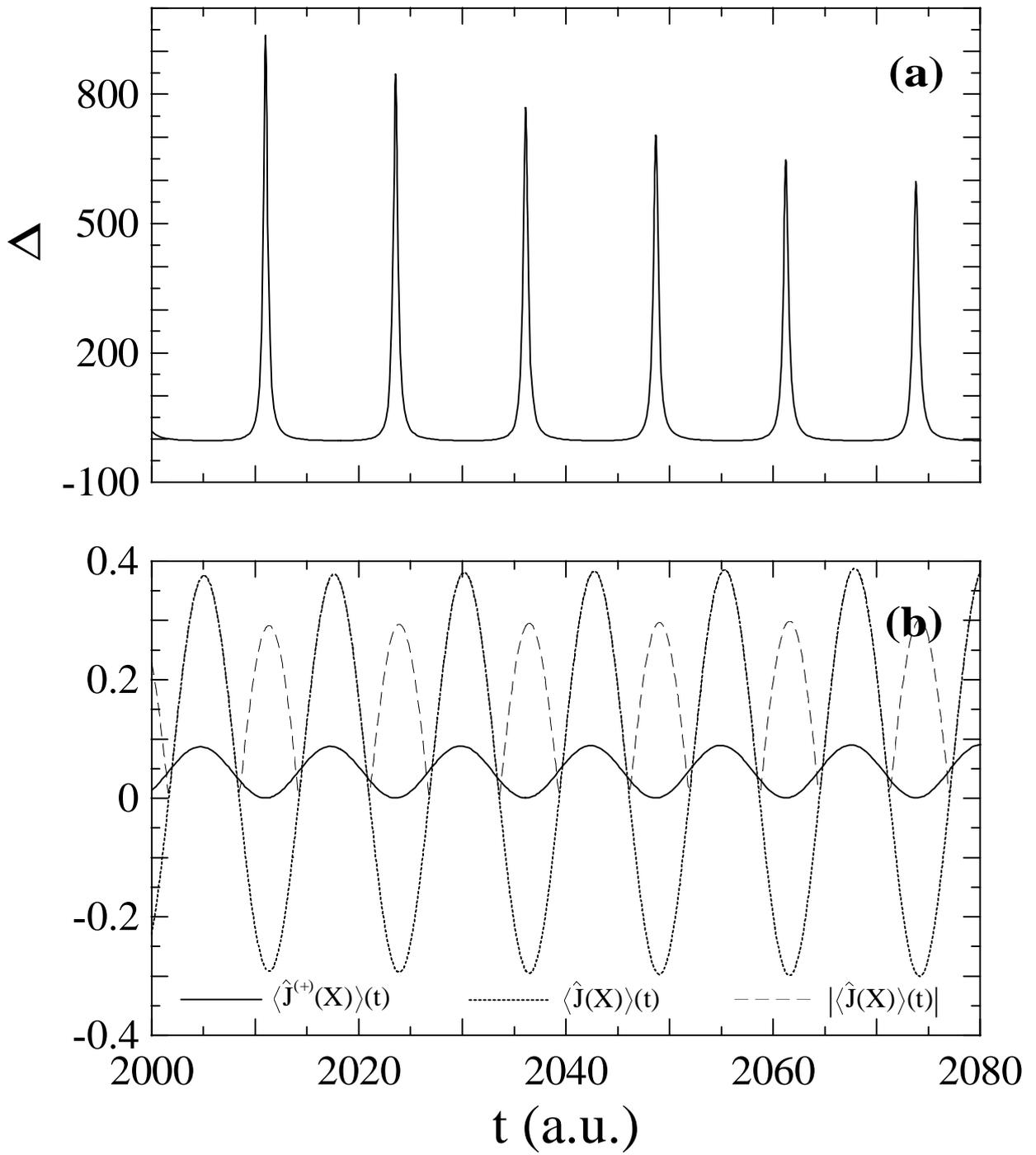
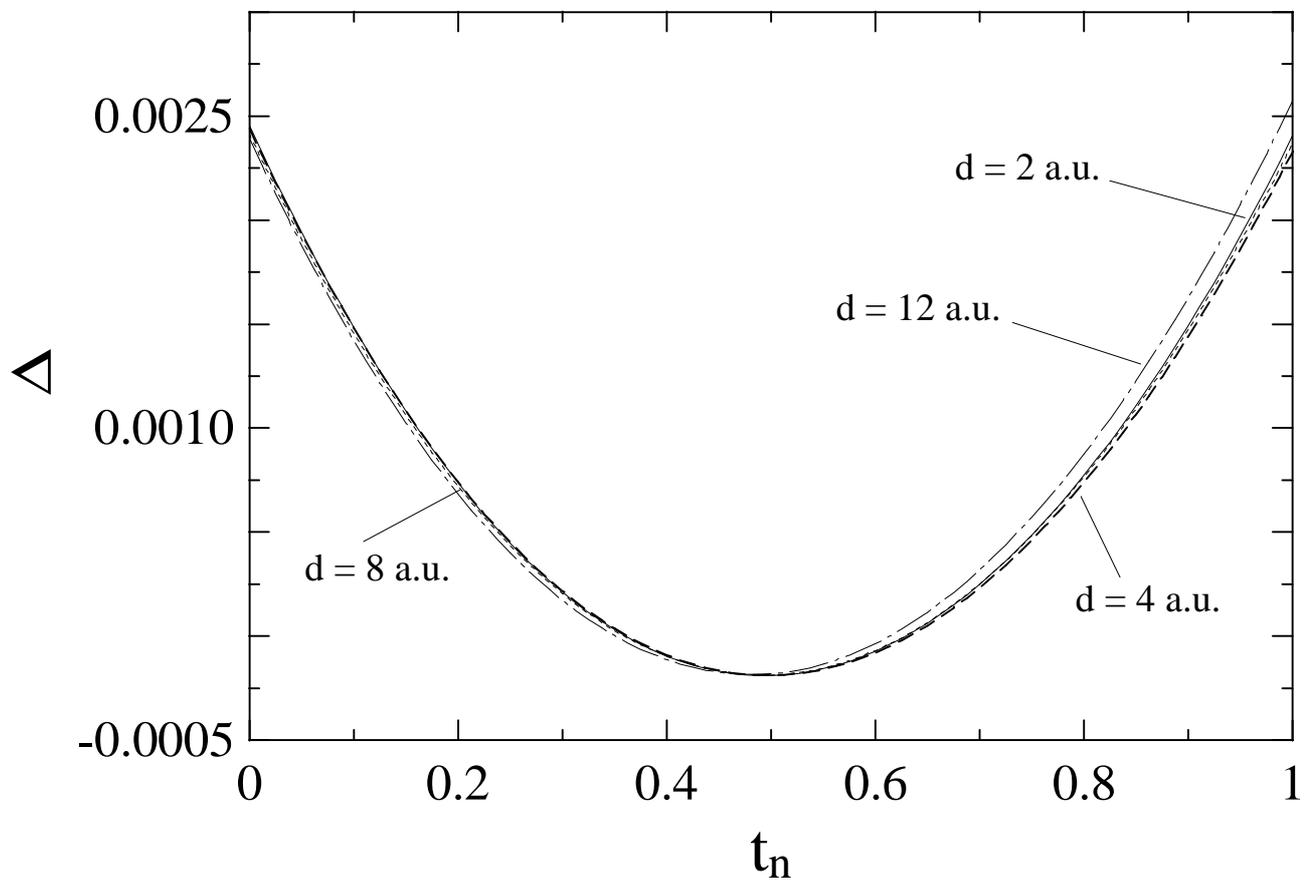
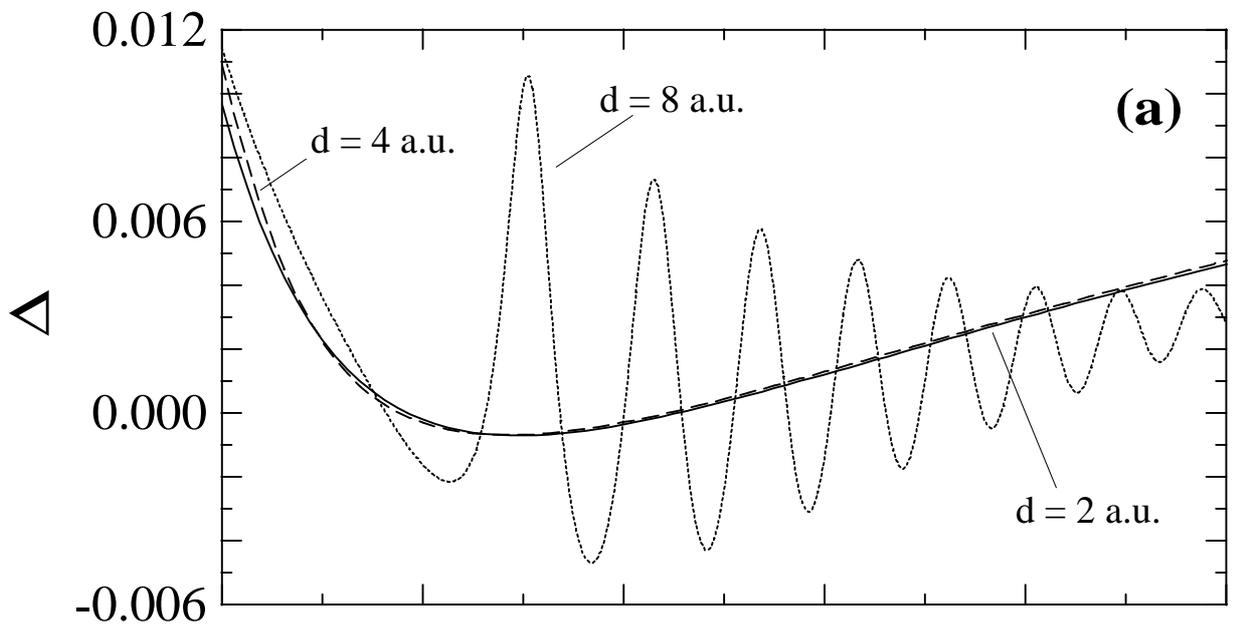
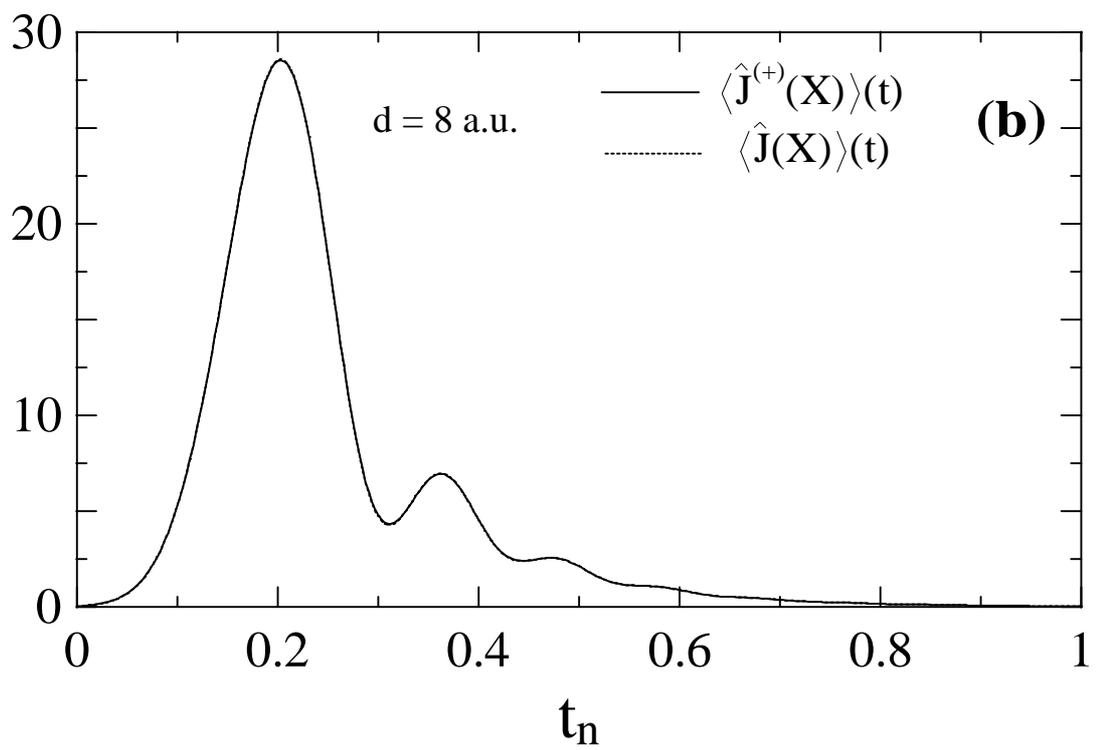
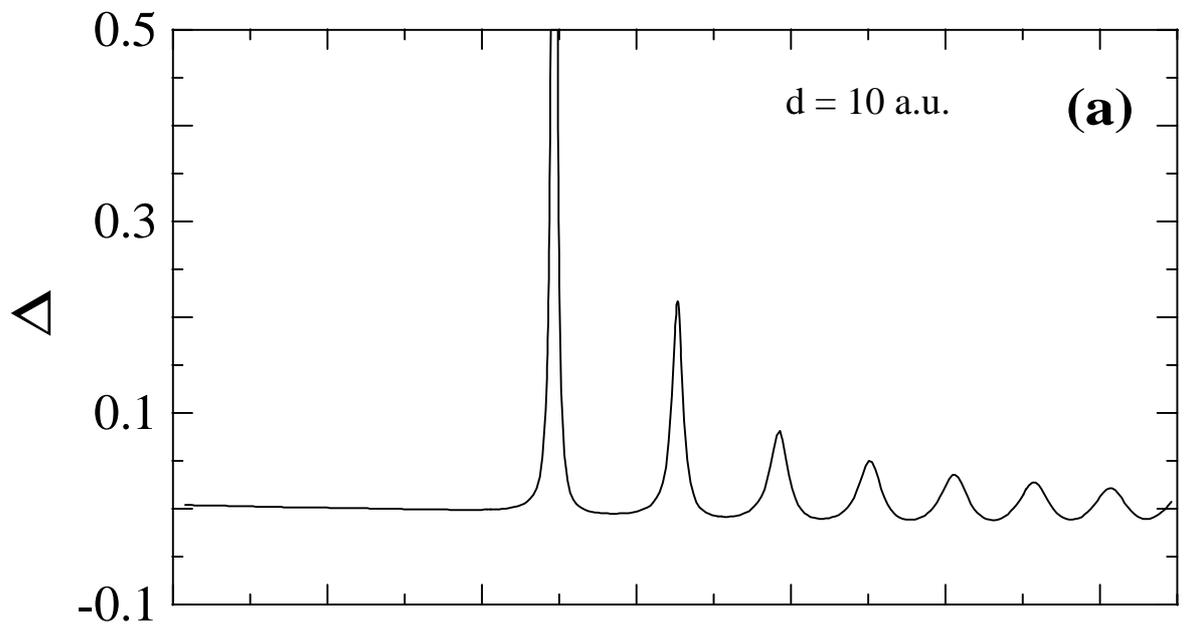
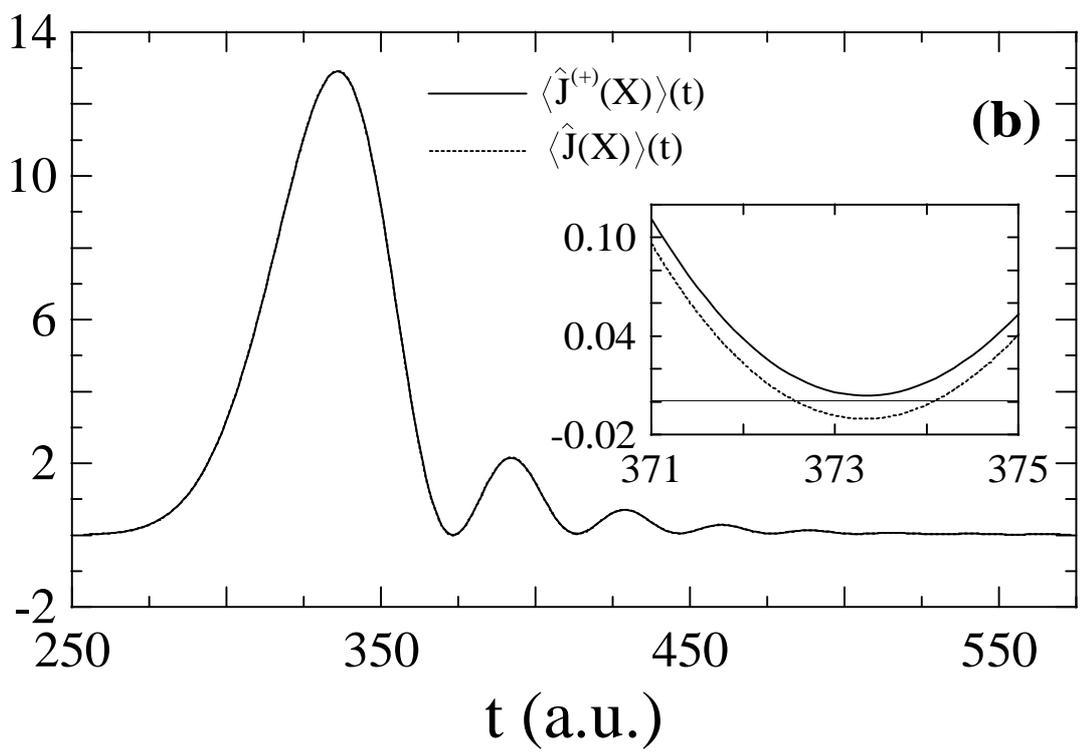
\end{document}